\documentclass[reprint,
superscriptaddress,
amsmath,
amssymb,
aps]{revtex4-2}

\usepackage{graphicx}% Include figure files
\usepackage{dcolumn}% Align table columns on decimal point
\usepackage{bm}% bold math
\usepackage{hyperref}% add hypertext capabilities
\usepackage{xcolor}% text color
\usepackage{breqn}
\usepackage{cancel}
\usepackage[normalem]{ulem}
\usepackage{setspace}
\usepackage[titletoc]{appendix}
\usepackage{ulem}

\makeatletter
\let\cat@comma@active\@empty
\makeatother

\begin{document}
\makeatletter
\gdef\@ptsize{1} % 1 for 11pt doc, 2 for 12pt
\makeatother
\preprint{APS/123-QED}

\title{ {Adaptive Optical Imaging with Entangled Photons}}

\author{Patrick Cameron} \email[Corresponding author: ]{p.cameron.1@research.gla.ac.uk}
\affiliation{School of Physics and Astronomy,University of Glasgow, Glasgow G12 8QQ, UK\
}%
\author{Baptiste Courme}
\affiliation{Sorbonne Université, CNRS, Institut des NanoSciences de Paris, INSP, F-75005 Paris, France\
}
\affiliation{Laboratoire Kastler Brossel, ENS-Universite PSL, CNRS, Sorbonne Universite, College de France, 24 rue Lhomond, 75005 Paris, France\
}
\author{ {Chloé Vernière}}
\affiliation{Sorbonne Université, CNRS, Institut des NanoSciences de Paris, INSP, F-75005 Paris, France\
}

\author{{Raj Pandya}}
\affiliation{Sorbonne Université, CNRS, Institut des NanoSciences de Paris, INSP, F-75005 Paris, France\
}
\affiliation{Laboratoire Kastler Brossel, ENS-Universite PSL, CNRS, Sorbonne Universite, College de France, 24 rue Lhomond, 75005 Paris, France\
}%
\affiliation{Cavendish Laboratory, University of Cambridge,
JJ Thomson Avenue, Cambridge, CB3 0HE, UK\
}%
\author{Daniele Faccio}%
\affiliation{School of Physics and Astronomy,University of Glasgow, Glasgow G12 8QQ, UK\
}%
\author{Hugo Defienne} \email[Corresponding author: ]{hugo.defienne@insp.upmc.fr}
\affiliation{School of Physics and Astronomy,University of Glasgow, Glasgow G12 8QQ, UK\
}%
\affiliation{Sorbonne Université, CNRS, Institut des NanoSciences de Paris, INSP, F-75005 Paris, France\
}


\begin{abstract}
Adaptive optics (AO) has revolutionized imaging in {fields} from astronomy to microscopy by {correcting} optical aberrations. In label-free microscopes, however, conventional AO {faces limitations} due to the absence of guidestar and the need to select an optimization metric specific to the sample and imaging process. Here, we propose an AO approach {leveraging} correlations between entangled photons to directly correct the point spread function (PSF). This guidestar-free method is independent of the specimen and imaging modality. We demonstrate the imaging of biological samples in the presence of aberrations using a bright-field imaging setup operating with a source of spatially-entangled photon pairs. Our approach performs better than conventional AO in correcting {specific} aberrations, particularly {those} involving significant defocus. Our work improves AO for label-free microscopy and could play a major role in the development of quantum microscopes.

\end{abstract}

\maketitle 

\section{Introduction}

Label-free microscopes are essential for studying biological systems in their most native states. In recent years, their performance {has been enhanced} by the use of non-classical light sources. In particular, sources of entangled photon pairs, which illuminate an object and are detected in coincidence to form an image, are at the basis of numerous protocols~\cite{moreau_imaging_2019}. For example, they are used in bright-field imaging configurations to enhance the spatial resolution~\cite{boto_quantum_2000,toninelli_resolutionenhanced_2019,defienne_pixel_2022,he_quantum_2023}, achieve sub-shot-noise imaging~\cite{brida_experimental_2010} and improve the contrast in the presence of noise and losses~\cite{defienne_quantum_2019,gregory_imaging_2020}. In phase imaging, they can be utilized to augment the contrast in both confocal~\cite{ono_entanglementenhanced_2013} and wide-field~\cite{camphausen_quantumenhanced_2021,black_quantumenhanced_2023} differential interference contrast (DIC) systems, and are at the basis of new modalities including quantum holography~\cite{defienne_polarization_2021,topfer_quantum_2022}, reconfigurable phase-contrast microscopy~\cite{hodgson_reconfigurable_2023} and 3D-imaging~\cite{zhang_ray_2022}. Finally, they can also improve time-gated imaging protocols, such as optical coherence tomography (OCT), by reducing dispersion~\cite{nasr_demonstration_2003,abouraddy_quantumoptical_2002} and enhancing depth sensitivity~\cite{ndagano_quantum_2022}. However, whether in their classical or quantum version, all these methods are sensitive to optical aberrations, created by the specimens being imaged or the imaging system itself. If left uncorrected, these effects negate the benefits gained by these techniques and compromise their practical use. 

Adaptive optics (AO) can be used to mitigate these aberrations. To operate, a light-emitting source or a point-like structure in the sample is identified as a guide star. The wavefront accumulates aberrations while propagating out of the sample, which are then measured by a Shack-Hartmann sensor (direct AO) or a focus-forming process (indirect AO). Wavefront correction is then applied to cancel out the aberrations using a deformable mirror or a spatial light modulator (SLM). Over the past decades, AO has played a major role in the development of advanced imaging systems, particularly fluorescence microscopes~\cite{ji_adaptive_2017,zhang_adaptive_2023}. 

In the absence of guide star, however, the point spread function (PSF) and thus the aberration information {is} not directly accessible. This is especially the case in most label-free and linear microscopy systems. To circumvent this issue, {wavefront sensorless, image-based AO methods} have been developed~\cite{zhang_adaptive_2023,hampson_sensorless_2020,booth_wave_2006}. They are based on the principle that the image, resulting from the convolution between the specimen and the PSF, has optimum quality only when the aberrations have been fully compensated. In practice, an image metric is first defined and then optimised by acting {with} the wavefront shaping device. The appropriate choice of the metric depends on the image formation process of the microscope used and the nature of the sample. The most commonly used include the total output intensity~\cite{debarre_imagebased_2009}, image contrast~\cite{zhou_contrastbased_2015}, low frequency content~\cite{debarre_image_2007} and sharpness~\cite{fienup_aberration_2003,murray_wavefront_2005}. In recent years, this has enabled aberration correction in several label-free microscope modalities, such as bright-field~\cite{debarre_image_2007}, quantitative phase-contrast~\cite{shu_adaptive_2022}, DIC~\cite{kam_computational_2001} and OCT~\cite{jian_wavefront_2014}. 
\begin{figure*}[ht]
    \centering
    \includegraphics[width = 0.9\textwidth]{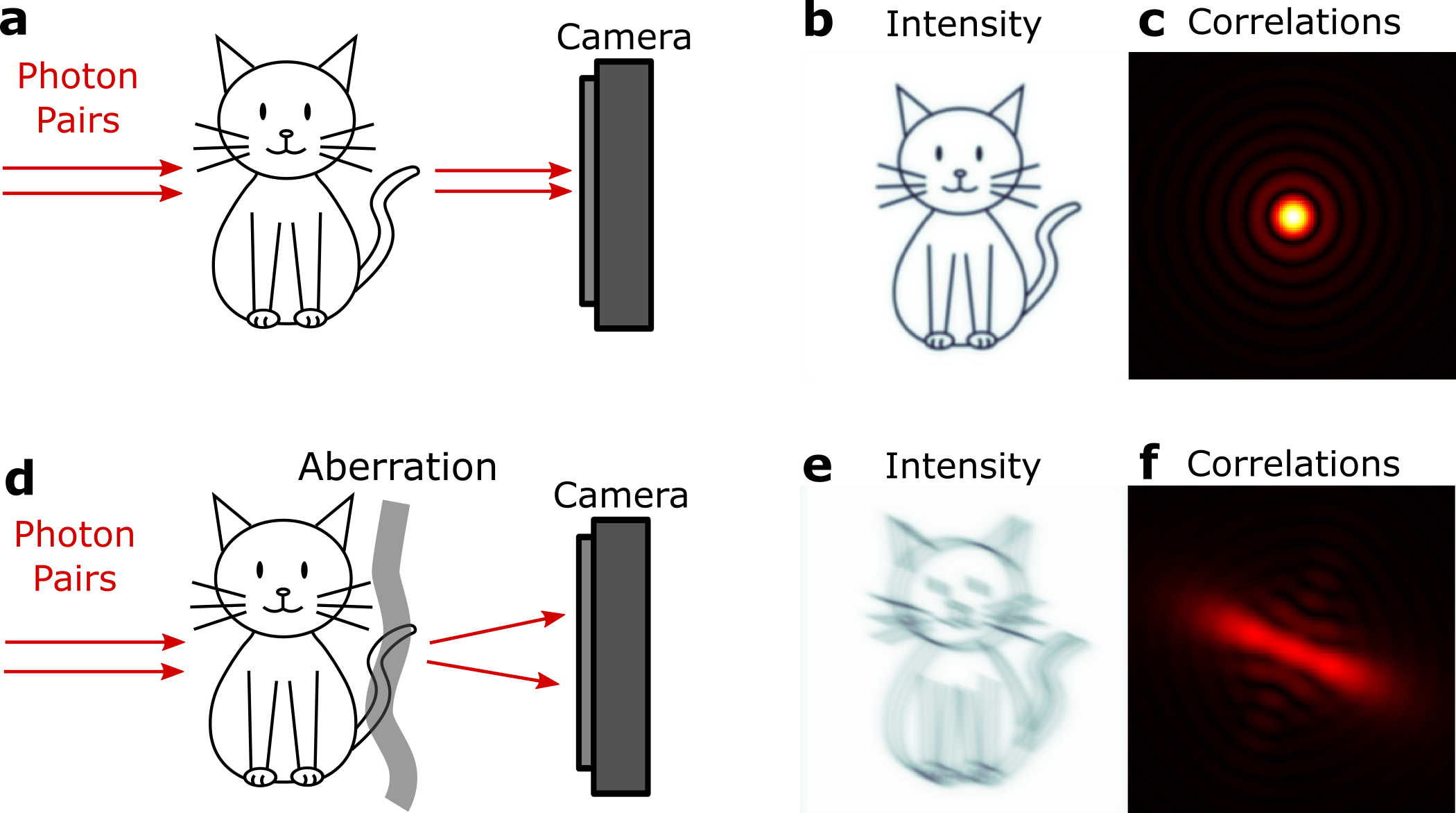}
    \caption{\textbf{Concept of {Quantum-assisted} Adaptive Optics {(QAO)}.} \textbf{a, } An object is illuminated by spatially entangled photon pairs and imaged onto a single-photon sensitive camera. The imaging system between the object and the camera is not represented for clarity. Photon pairs are strongly correlated in the object plane. Without optical aberrations, a \textbf{(b)} sharp intensity image of the object is acquired and photon pairs are still correlated at the camera plane. Photon pairs correlations are visualized by \textbf{(c)} measuring the spatial second-order correlation function, $G^{(2)}$, and projecting it onto specific coordinates. Such a  $G^{(2)}$ projection is proportional the system's point-spread function (PSF) and shows a narrow peak at its center. \textbf{d}, With aberrations present, the system is not limited by diffraction and the pairs are no longer correlated at the camera plane, resulting in a \textbf{(e)} blurred intensity image and a \textbf{(f)} distorted $G^{(2)}$ projection. In QAO, aberrations are corrected using a spatial light modulator (SLM) to maximize the central value of the $G^{(2)}$ projection.} 
    \label{fig:concept}
\end{figure*}

One of the primary hurdles in achieving effective image-based AO lies in the requirement to define distinct metrics for each microscope modality and for varying specimen types. Furthermore, certain metrics may introduce systematic errors. For instance, when capturing volumetric samples, the utilization of an image sharpness metric to correct defocus aberration typically yields multiple solutions corresponding to different imaging planes within the sample. 

In this work, we present a quantum-assisted AO (QAO) method that harnesses {the} entanglement between photon pairs to directly access the imaging system PSF, and thus the aberration information, in the absence of a guidestar. This approach also eliminates the need to define a specific image-based metric, and is thus independent of the imaging modality and specimen under study. We demonstrate the effectiveness of this approach by imaging biological samples using classical and quantum bright-field transmission imaging systems in the presence of aberrations. In particular, we show experimental situations in which it leads unambiguously to the optimal correction, while classical image-based AO methods fail.

\section{Concept}

In QAO scheme, spatially-entangled photon pairs are incident on an object ($t$) which is then imaged onto a single-photon sensitive camera (Fig.~\ref{fig:concept}.a). As in classical incoherent illumination, {the} intensity image ($I$) produced at the output results from a convolution between the absolute value-squared PSF ($h$) and the object as $I = |h|^2 * |t|^2$ (Fig.~\ref{fig:concept}.b). In addition, the photons forming the image are also pairwise correlated in space, which arises from their entanglement~\cite{walborn_spatial_2010}. The second-order spatial correlation function $G^{(2)}$ can be written as
\begin{equation}
    G^{(2)}(\mathbf{r_1',r_2'}) = \left | \phi(\mathbf{r_1},\mathbf{r_2}) t(\mathbf{r_1}) t(\mathbf{r_2}) * h(\mathbf{r_1}) h(\mathbf{r_2}) \right |^2,
    \label{Eq1}
\end{equation}   
where $\phi(\mathbf{r_1},\mathbf{r_2})$ is the spatial two-photon wavefunction of the photon pair in the object plane with transverse coordinates $\mathbf{r_1}$ and $\mathbf{r_2}$~\cite{abouraddy_entangledphoton_2002}. In general, $G^{(2)}$ is a {complicated} function that depends on the PSF, the object and the spatial correlations between photon pairs. 
Under specific experimental conditions, however, one can simplify Equation~\ref{Eq1} and average $G^{(2)}$ along specific spatial axes to extract information only linked to the system's PSF. In particular, if the object is positioned in the Fourier plane of the source, the two-photon wavefunction can be approximated by $\phi(\mathbf{r_1},\mathbf{r_2}) \approx  \delta(\mathbf{r_1}+\mathbf{r_2})$, which describes near-perfect {anti-correlations between photon pairs originating from spatial entanglement.}. {In this configuration we can measure the sum coordinate projection of $G^{(2)}$, defined as $C^{+}(\bm{\delta r^{\scriptstyle +}}) = \int G^{(2)}(\mathbf{r},\bm{\delta r^{\scriptstyle +}}-\mathbf{r}) d\mathbf{r}$, with $\bm{\delta r^{\scriptstyle +}} = \mathbf{r_1}+\mathbf{r_2}$ being the sum-coordinate. Assuming weak optical aberrations in the imaging system, $C^{+}$} can be approximated as:{
\begin{equation}
    C^{+}(\bm{\delta r^{\scriptstyle +}}) \approx K \left| [h*h] (\bm{\delta r^{\scriptstyle +}}) \right|^2, 
    \label{eq:cplus}
\end{equation}
     
where $K{= \int |t(\mathbf{r}) t(-\mathbf{r})|^2 d(\mathbf{r})}$ is a constant independent of $h$. {$K$ represents the  photon-pair transmission rate through the sample.} For example, Figure~\ref{fig:concept}.c shows a sum-coordinate projection simulated in the case of a diffraction-limited imaging system. It has a very specific shape, with a narrow peak at its center, just like the corresponding PSF. In the presence of optical aberrations, however, the PSF is distorted, as is the sum-coordinate projection (Fig.~\ref{fig:concept}.f), with a central correlation peak that decreases and spreads. {The value of the central peak is therefore}  maximal when the imaging system is limited by diffraction. In QAO, we use this value as a feedback signal to compensate for optical aberrations in the imaging system using a modal-based adaptive optics algorithm. Simulations and additional experimental data supporting this result are provided in section {III} of the supplementary document. 

\begin{figure*}[ht]
    \centering
    \includegraphics[width = \textwidth]{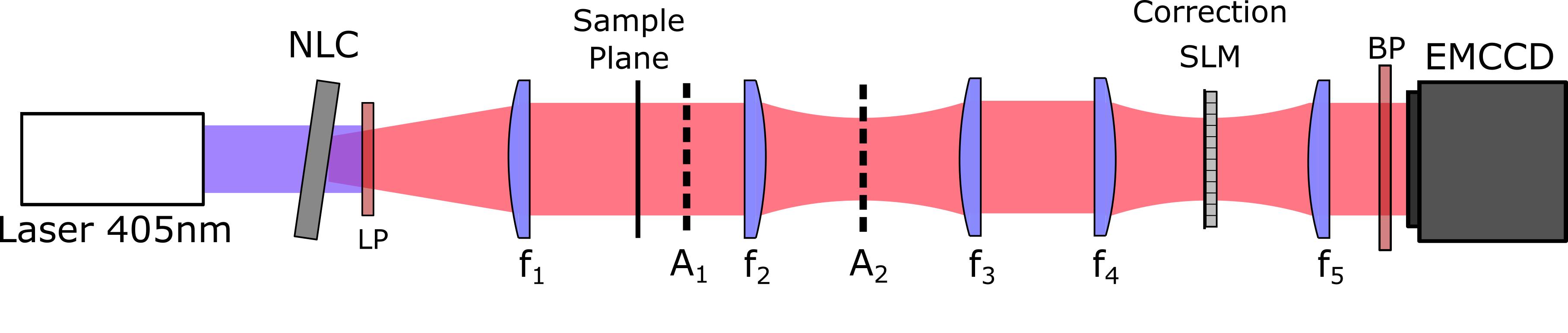}
    \caption{\textbf{Experimental setup.} Spatially-entangled photon pairs centered at 810nm are produced via Type I Spontaneous Parametric Down-Conversion (SPDC) using a 405nm collimated continuous-wave laser and a 0.5mm-thick $\beta$-Barium Borate nonlinear crystal (NLC).
     Blue photons are then filtered out by a low-pass filter (LP) at 650nm. The sample is illuminated by the photon pairs while being positioned in the Fourier plane of the crystal ($f_1=100mm$). It is subsequently imaged (with a magnification of 1) onto the electron-multiplying charge-coupled device (EMCCD) camera using two 4-f imaging systems, $f_2-f_3$ and $f_4-f_5$. The spatial light modulator (SLM) used to correct aberrations is positioned in a Fourier plane of the sample between $f_4$ and $f_5$. For clarity, it is depicted in transmission, but in practice, it operates in reflection. Optical aberrations can be introduced at either the optical planes A1 (near the sample plane) or A2 (Fourier plane). {Note that plane A1 is deliberately placed at a small distance from the object plane to introduce sufficient aberrations.} To detect only near-degenerate photon pairs, a bandpass filter (BP) at $810\pm10$nm is positioned in front of the camera.}
    \label{fig:setup}
\end{figure*}

\section{Results}

Figure~\ref{fig:setup} shows the experimental setup. Spatially entangled photon pairs are generated via spontaneous parametric down conversion (SPDC) in a thin $\beta$-barium-borate (BBO) crystal cut for Type I phase matching. Using lens $f_1$, the output surface of the crystal is Fourier-imaged onto the sample. Subsequently, the sample is imaged onto the camera using two $4f$ imaging systems, namely $f_2-f_3$ and $f_4-f_5$. Specimen- and system-induced aberrations can be introduced in the imaging system in planes A1 and A2, respectively. A spatial light modulator (SLM), used to correct for aberrations, is placed in a Fourier plane of the sample. Photon pairs transmitted through the system are detected at the output using an electron multiplying charge coupled device (EMCCD) camera. This is utilized for measuring both conventional intensity images and photon correlations, following the technique described in Ref.~\cite{defienne_general_2018} {( see also sections I and II of the supplementary document)}.

To illustrate our approach, we place a biological sample - a honeybee mouthpiece on a microscope slide - in the sample plane and capture its intensity image in transmission (Fig.~\ref{fig:results}a). In the absence of aberrations, the sum-coordinate projection exhibits a distinct and sharp peak, as shown in Figure~\ref{fig:results}d. However, when aberrations are present, the image becomes blurred, and the correlation peak is spread and distorted, as depicted in Figures~\ref{fig:results}b and {~\ref{fig:results}}e, respectively. In this demonstration, we induce aberrations by introducing a second SLM at plane A2 that displays a low-frequency random phase pattern (see {section V of the supplementary document}). 

To correct aberrations, we employ a modal-based adaptive optics algorithm that utilizes $C^+_0$ as a feedback parameter{, where $C_0^+=C^+(\bm{\delta r^+=0})$}. This algorithm involves introducing predetermined aberrations on the SLM using Zernike polynomial modes. In our study, we consider all modes with radial numbers $n \leq 5$ and azimuthal numbers $|m| \leq n$, excluding piston, tip, and tilt. For each Zernike mode ($Z^m_n$), we record five sum-coordinate projections with distinct, known bias amplitudes ($\alpha_{nm}$). In each measurement, the SLM phase $\theta_{nm}$ is thus modulated according to the Equation: 
\begin{equation}
    \theta_{nm} = \theta_{nm-1} + \alpha_{mn} Z^m_n,
\end{equation}
where $\theta_{nm-1}$ represents the optimal phase correction obtained for the previous mode. {Such a phase modulation approach is commonly used in classical modal AO~\cite{booth_wave_2006}}.
For example, the values of $C^+_0$ obtained from the sum-coordinate projections for the modes $Z_{3}^{-3}$ and $Z_3^{1}$ are shown in Figure~\ref{fig:results}h. The positions of the maxima, denoted $\alpha_{-33}^{corr}$ and $\alpha_{13}^{corr}$, representing the optimal corrections for their respective mode, are determined using a Gaussian fitting model (see section IV of the supplementary document). After several optimization steps, a narrow peak is recovered in the sum-coordinate projection (Fig.~\ref{fig:results}f) and a sharp image appears in the intensity (Fig.~\ref{fig:results}c). Visual comparison with the aberration-free images shows a clear improvement after correction. Quantitatively, one can use the structural similarity (SSIM) as a metric to quantify image similarity. Using the aberration-free image as a reference (Fig.~\ref{fig:results}a), we find $SSIM=77.89 \%$ for the uncorrected image (Fig.~\ref{fig:results}c) and $SSIM = 98.41\%$ for the corrected image (Fig.~\ref{fig:results}b). Note that here, although the object is illuminated by a source of entangled photon pairs, whose quantum properties are crucial for measuring $C^+$ and thus correcting aberrations, the imaging process itself is purely `classical'  as the output image is obtained through a simple intensity measurement.

\begin{figure*}
    \centering
    \includegraphics[width = 0.8\textwidth]{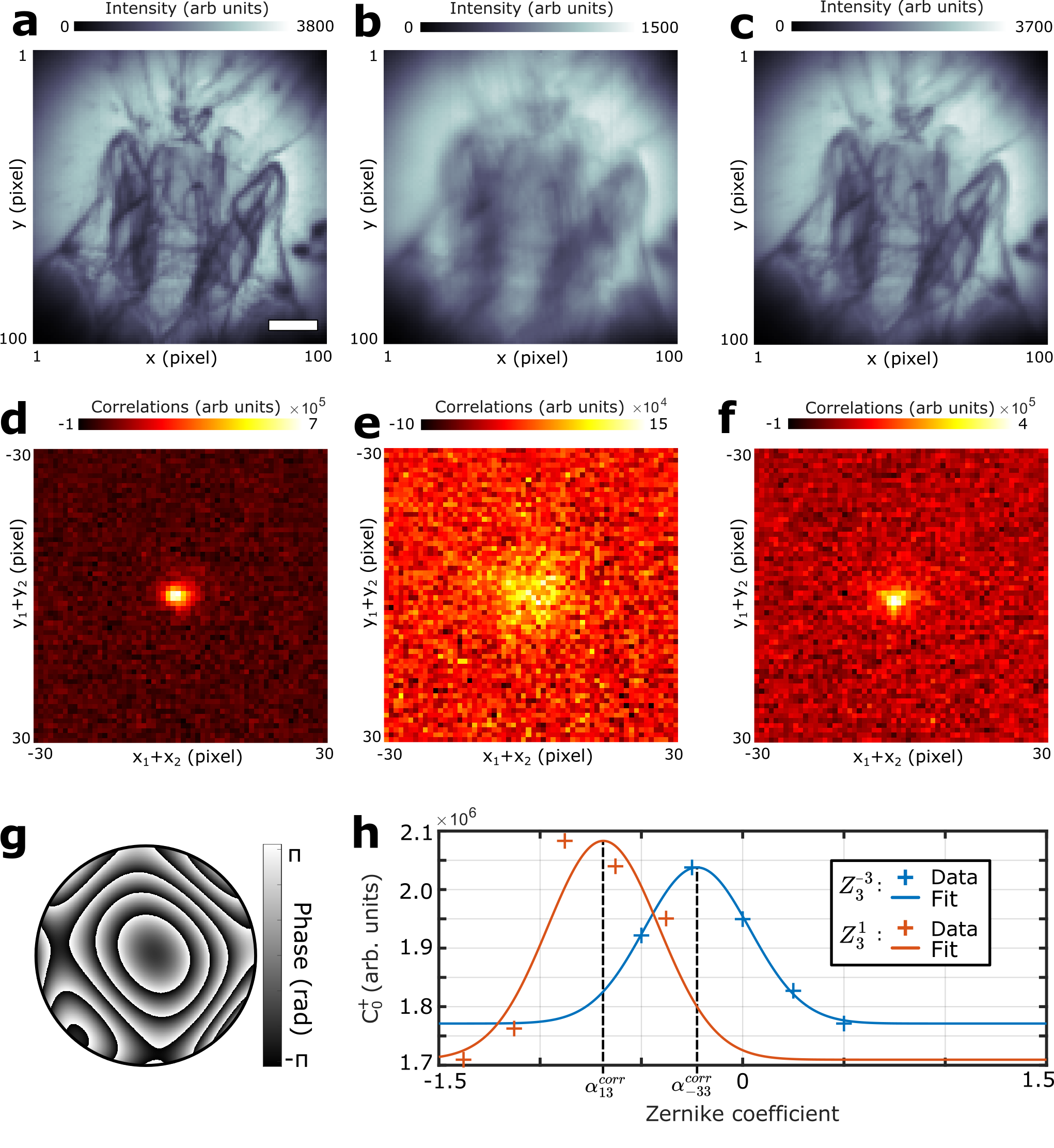}
    \caption{\textbf{Results of QAO correction.} \textbf{a-c}, Intensity images of a biological sample (bee head) acquired in transmission \textbf{(a)} without aberrations, \textbf{(b)} with aberrations before correction, and \textbf{(c)} after correction.  Using the aberration-free intensity image as a reference, we find structural similarity values of $SSIM=77.89\%$ and $SSIM=98.41\%$ for the uncorrected image and the corrected image, respectively. \textbf{d-f}, Correlations images $C^+(\bm{\delta r^{\scriptstyle +}} = \mathbf{r_1}+\mathbf{r_2})$ measured \textbf{(d)} without aberrations, \textbf{(e)} with aberrations before correction, and \textbf{(f)} after correction. \textbf{g,} Optimal phase pattern obtained after correction and displayed on the SLM. \textbf{h, }  Values of the sum-coordinate projection peaks $C_0^+$ in function of the coefficient $\alpha_{mn}$ for two Zernike modes $Z_3^{-3}$ and $Z_3^{1}$ (crosses). $\alpha^{corr}_{-33}=-0.2253$ and $\alpha^{corr}_{13}=0.6881$ are the two optimal correction values for each mode returned by the fit (solid lines). Each intensity and sum-coordinate projection was obtained from $10^5$ frames, approximately equivalent to a $2$ min-acquisition. The white scale bar {is} 400$\mu$m.} 
    \label{fig:results}
\end{figure*}

QAO offers several advantages compared to classical AO. Firstly, as demonstrated in Figure~\ref{fig:results}, it does not require a guidestar. All photon pairs forming the image possess information about the system aberrations at every point, because these are encoded in their spatial correlations. Additionally, QAO performance does not depend on the sample properties or the imaging modality. The spatial correlation structure is a property of the illumination itself, and is only affected by the system aberrations. This implies that QAO will converge irrespective of the observed sample type, ranging from nearly transparent samples (e.g. cells) to denser ones (e.g. layered minerals), regardless of their complexity or smoothness of structure. In {Figure S9} of the supplementary document, we provide additional experimental results obtained using various sample types {that demonstrate this}. In this aspect, QAO thus surpasses all image-based AO techniques, where the chosen metrics and optimization performances depend on the properties of the sample. More remarkably, we show in the next section that, in certain imaging situations, image-based approaches can lead to systematic error in aberration correction, whereas QAO converges to the correct solution.

We consider a situation where the sample has a 3-dimensional structure, which is very common in microscopy. In such a case, it is known that it is not possible to correct for defocus aberrations properly. Indeed, when using an image quality metric, it may optimize for the wrong focal plane within the sample. Since the sample structure has no effect in QAO, defocus correction is possible. In our demonstration, we chose an object consisting of three copper wires, each with an approximate thickness of $0.15$ mm, and spaced approximately $5$ mm apart along the optical axis. We then induced defocus aberration with strength $\alpha^{aber}_{02}=-2$ by placing a second SLM in plane A2. Sum-coordinate projection and intensity images are acquired for a wide range of defocus corrections ($\alpha_{02} \in [-5,5]$) programmed on the correction SLM. At each step, values of three standard AO image quality metrics are calculated from the intensity image: power in bucket (PIB)~\cite{debarre_imagebased_2009}, image contrast~\cite{zhou_contrastbased_2015}, and low frequency content~\cite{debarre_image_2007}. In addition, $C_0^+$ is also retrieved from the sum-coordinate projection. Figure~\ref{fig:3 wires}.a shows the four corresponding optimization curves. First, we observe that the various classical AO metrics return different optimization coefficients, highlighting their dependency on the object's structure. 
Then, by examining the intensity images captured while programming each optimal correction phase pattern (Figs.~\ref{fig:3 wires}b-e), it becomes evident that none of these metrics properly corrected the aberrations. Indeed, the aberration-free image in Figure~\ref{fig:3 wires}g clearly shows that only the bottom wire is in the focal plane, which is not the case in any of the intensity images shown in Figures~\ref{fig:3 wires}b-e. On the other hand, QAO converges to the correct solution, as seen in the intensity image shown in Figure~\ref{fig:3 wires}f ($SSIM = 96.83\%$). Interestingly, we also note that the optimum value found with QAO is $\alpha^{corr}_{02} = 1.622$, which differs slightly from the value of $2$ (opposite of $\alpha^{aber}_{02}=-2$) that we would be expect to find. This is because QAO corrects not only for the intentionally introduced defocus aberrations in the A2 plane, but also for those inherent in the imaging system. This is also shown by the fact that the correlation peak in Figure~\ref{fig:3 wires}.f (inset) is slightly narrower than the one in Figure~\ref{fig:3 wires}.g (inset). {This demonstration uses a very simple three-dimensional sample: three spaced wires. However, QAO can in principle be used with more complex three-dimensional samples as long as we remain within the regime of weak aberrations, i.e., no strong scattering and absorption. Such samples are typically studied with optical tomography methods, where QAO can therefore be used after adapting the mathematical formalism to account for the thickness of these objects~\cite{zhang_ray_2022}.}   

\begin{figure*}
    \centering
    \includegraphics[width = 0.7\textwidth]{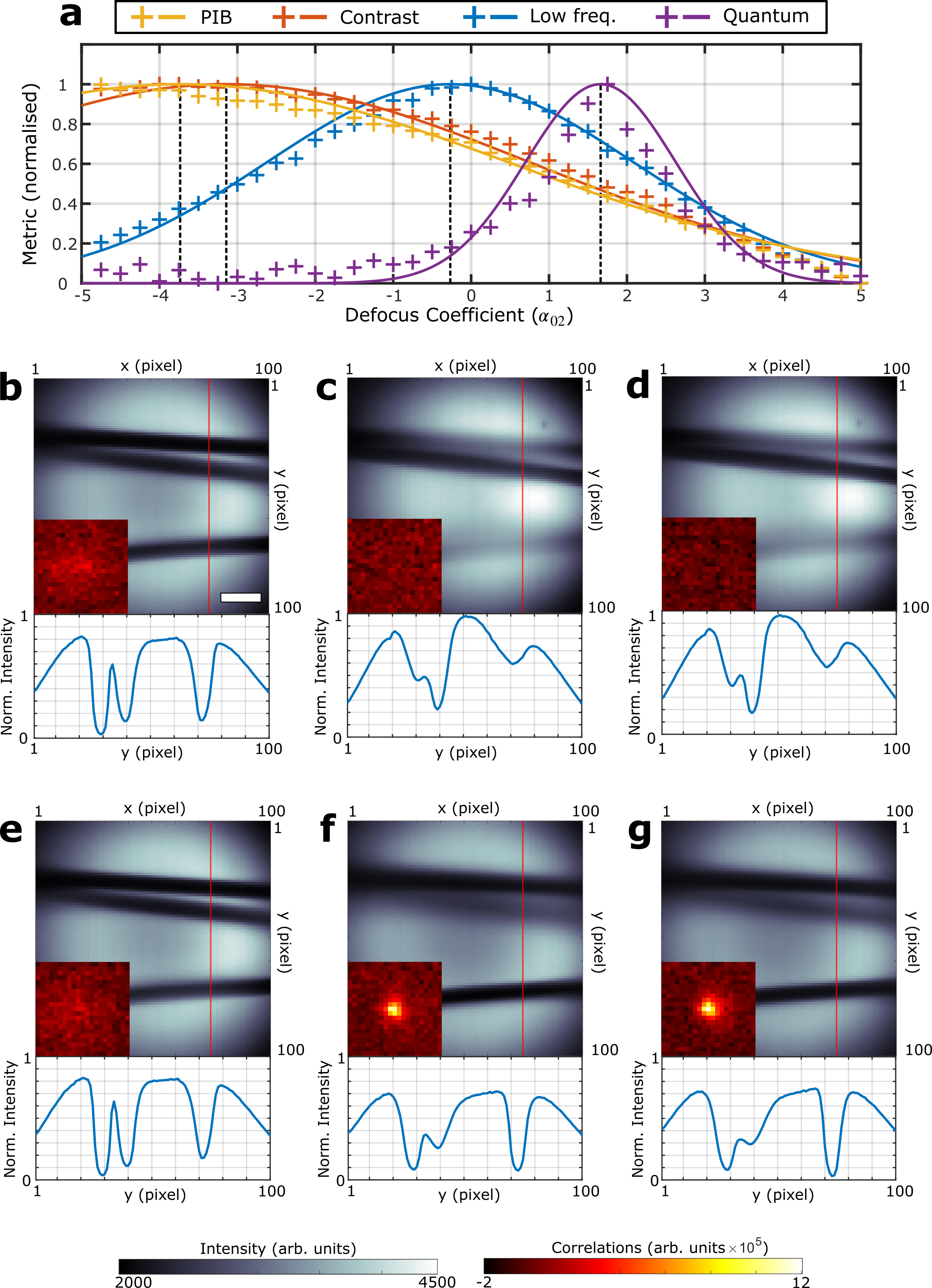}
    \caption{\textbf{Comparison between QAO and classical image-based AO}. \textbf{a, } Values of three image quality metrics (namely power in bucket (PIB), image contrast and low frequencies) and $C_0^+$ in function of the defocus correction coefficient $\alpha_{02}$. Data is given by the crosses, and the fits used to find optimal values ($\alpha^{corr}_{02}$) are given by solid lines. In this experiment, the object is 3-dimensional (three thin copper wires). \textbf{ b-g,} Intensity images (grayscale), central regions of $C^+$ (inset) and intensity profile for a single column (line plot) for various defocus corrections on SLM: \textbf{b,} without correction ($\alpha^{corr}_{02} = 0 $ and $SSIM = 76.39 \%$); \textbf{c,} Optimal correction found using a `Power in Bucket' metric ($\alpha^{corr}_{02} = -3.1427$ and $SSIM = 50.56\%$); \textbf{d,} Optimal correction found using a `Image Contrast' metric ($\alpha^{corr}_{02} = -3.1427$ and $SSIM = 52.29\%$); \textbf{e,} Optimal correction found using a `Low Spatial Frequencies' metric ($\alpha^{corr}_{02} = -0.2677$ and $SSIM = 72.61\%$); \textbf{f,} Optimal correction found using QAO ($\alpha^{corr}_{02} = 1.6622$ and $SSIM = 96.83\%$); \textbf{g,} No aberration. Vertical red lines show selected column for profile plots. Each intensity image and sum-coordinate projection were obtained from $10^5$ frames, approximately equivalent to a $2$ min-acquisition. The white scale bar {is} 400$\mu$m.}
    \label{fig:3 wires}
\end{figure*}

Finally, in order to showcase its potential for quantum imaging, QAO is applied to a `quantum' variant of the bright-field imaging setup depicted in Figure~\ref{fig:setup}. In such a scheme, only one photon of a pair interacts with the object, while its twin serves as a reference. For that, the sample is placed on only one half of the object plane ($x>0$), as observed in the intensity images shown in Figures~\ref{Figure5}.a and b. {To interpret this specific arrangement in Equation~\ref{Eq1}, we theoretically define the object such that $t(x<0)=1$ and $t(x>0)$ describes the object.} Then, the final image ($R$) is obtained by measuring photon correlations between all symmetric pixel pairs of the two halves, i.e. $R(\mathbf{r}) \approx G^{(2)}(\mathbf{r},-\mathbf{r})$ (see section II of the supplementary document). This image is called an anti-correlation image and is shown in Figure~\ref{Figure5}.e. As demonstrated in previous studies~\cite{he_quantum_2023,defienne_polarization_2021, gregory_imaging_2020}, such a quantum scheme offers some advantages over its classical counterpart, including an enhanced transverse spatial resolution and increased resilience against noise and stray light. In the presence of aberrations, however, we show that this imaging technique becomes highly impractical and thereby loses all its purported advantages. For example, Figure~\ref{Figure5}.f shows an anti-correlation image acquired after inserting a 1-cm-thick layer of polydimethylsiloxane (PDMS, shown in Fig.~\ref{Figure5}.c) on both photon paths in plane A1 to induce optical aberrations. Not only is the resulting image blurred, leading to a complete loss of the expected resolution advantage, but also its signal-to-noise ratio (SNR) is greatly reduced, rendering the sample almost indiscernible ($SNR \approx 3$). After applying QAO, we retrieve an anti-correlation image in Figure~\ref{Figure5}.g that has a spatial resolution closer to that without aberrations and of much better quality ($SNR=15$). The inset of Figure~\ref{Figure5}.g shows the corresponding sum-coordinate projection, exhibiting a much narrower and {more} intense peak, and Figure~\ref{Figure5}.d shows the optimal SLM phase pattern. In addition, when comparing carefully the sum-coordinate projections without aberrations (inset in Figs.~\ref{Figure5}.e) and after correction (inset in Figs.~\ref{Figure5}.g), we observe that QAO also corrected for a small PSF asymmetry present in the initial system. Compensating for this asymmetry results in a more uniform output image (Fig.~\ref{Figure5}.g) than that obtained in the aberration-free case (Fig.~\ref{Figure5}.e). By using QAO, we then show a significant improvement of the output image quality in terms of resolution, SNR and uniformity, effectively restoring the operational capability of this quantum imaging technique.

\begin{figure*}
    \centering
    \includegraphics[width = 0.7\textwidth]{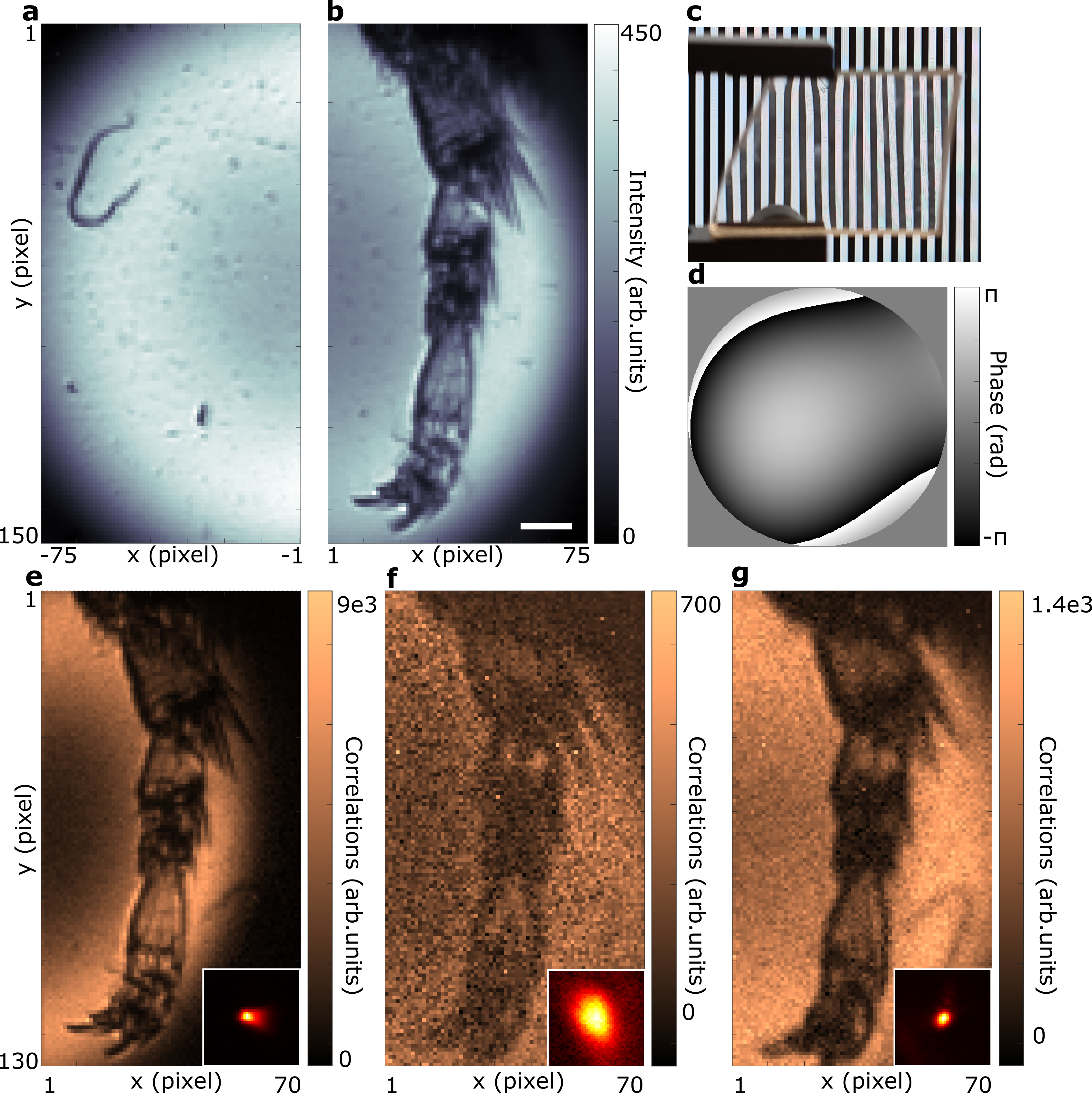}
    \caption{\textbf{Application to quantum imaging.} \textbf{a, } Intensity image formed by one photon of a pair used as a the reference photon. \textbf{b, } Intensity image formed by the other photon used to illuminate the sample, here a bee's leg. \textbf{c, } 1-cm-thick piece of polydimethylsiloxane (PDMS) inserted {in} plane A1 in the setup in Figure~\ref{fig:setup} to induce aberrations.  \textbf{d,} Optimal phase pattern obtained after correction and displayed on the SLM. \textbf{d-g,} Anti-correlation images $R(\mathbf{r}) \approx G^{(2)}(\mathbf{r},-\mathbf{r})$ obtained \textbf{(e)} without induced aberration (signal-to-noise ratio: $SNR=29$), \textbf{(f)} with aberration ($SNR \approx 3$), and \textbf{(g)} after aberration correction ($SNR=15$). Insets show the sum-coordinate projection in each case. Each sum-coordinate projection to achieve QAO was obtained from $2.2.10^4$ frames, approximately 3 mins of acquisition, and each anti-correlation image was obtained from $10^7$ frames, approximately $24$ hours of acquisition. Note that the EMCCD camera used here is different from this used in Figures~\ref{fig:results} and~\ref{fig:3 wires}, and has a frame rate of just 130 fps. The total intensity (i.e. total number of photons) measured on the camera was the same in the presence of aberrations before and after correction. The white scale bar represents 400$\mu$m. } 
    \label{Figure5}
\end{figure*}

\section{Discussion}

We have introduced a {Q}AO method that eliminates the need for a guidestar. By optimizing the spatial correlations of entangled photon pairs, we can directly optimise the system PSF and compensate for optical aberrations. QAO circumvents certain limitations linked to conventional image-based AO, and is particularly well-suited for classical and quantum full-field, label-free and linear microscopy systems. 

{In our study, we demonstrate QAO in the regime of weak optical aberrations. We use artificial layers to simulate aberrations commonly encountered in real-world microscopy systems, including system-induced (e.g. astigmatism, defocus, comatic aberrations due to objectives, and misalignment) and weak specimen-induced aberrations (e.g. translucent tissues surrounding the sample, immersion liquid, and sample support). At this stage, QAO is not demonstrated in the scattering regime, although preliminary results obtained with more complex aberrations show promise (supplementary Figures S14, S15 and S16). Within this regime of weak aberrations, there are no fundamental barriers preventing the use of QAO in other, more advanced label-free imaging systems. For instance, QAO could improve current image-based approaches used in optical coherence tomography~\cite{xiao_adaptive_2016}, be combined with 3D imaging techniques, some of which already utilized entangled photon pair sources~\cite{zhang_ray_2022}, be employed in phase imaging and high-numerical aperture imaging schemes (supplementary Figures S19 and S20) and adapted to reflection geometries by employing multiple SLMs (simulation in supplementary Figure S17). As with classical AO, the effectiveness of the correction found with QAO will always depend on the imaging modality and the nature of aberrations present. For instance, spatially variant aberrations will restrict the field of view in the corrected image, although this limitation might be circumvented by utilizing alternative AO designs like conjugate and multi-conjugate AO~\cite{mertz_field_2015,simmonds_modelling_2013}. Finally, it is important to note that QAO is not yet adaptable in fluorescence microscopy, but this could change in the future with the emergence of photon-pairs emitting biomarkers~\cite{frenkel_two_2023}.}

{In practice,} the main limitation of QAO is its long operating time. Using an EMCCD camera, acquisition times of the order of one minute are required to measure one sum-coordinate projection. This means that correcting for multiple orders of aberration can take up to several hours. 
However, this technical limitation can be overcome by employing alternative camera technologies, some of which are already available commercially.  For example, single-photon avalanche diode (SPAD) cameras have been employed to capture sum-coordinate projections at speeds up to 100 times faster than EMCCD cameras using similar photon pair sources~\cite{ndagano_imaging_2020,camphausen_fast_2023}. Another promising technology is the intensified Tpx3cam camera, which has recently been utilized for similar correlation measurements~\cite{nomerotski_intensified_2023,vidyapin_characterisation_2023,courme_quantifying_2023}. As technology improves, we thus expect acquisition times soon to be on the order of seconds, which would result in correction times on the order of minutes. In addition, here we chose Zernike polynomials as the basis set for aberration representation, even though they may not be optimal~\cite{hu_universal_2020}. In particular, if the aberrations are more complex, wavefront shaping approaches using Hadamard or random bases should be considered~\cite{yeminy_guidestarfree_2021,vellekoop_focusing_2007,defienne_adaptive_2018,lib_realtime_2020,courme_manipulation_2023}. 

{In our demonstration, QAO employs entanglement between photons. Indeed, replacing our source by classically-anti-correlated photons would yield a formally different output measurement i.e. $C_{cl}^+ = |h|^2 * |h|^2$ (supplementary section XIV). Such a metric could still be used for AO, but is genuinely less sensitive compared to entangled photons (supplementary Figure S18) and not suitable for phase-imaging. In addition, producing such near-perfect classical anti-correlations is challenging in practice. One potential approach could use thermal light, that is naturally position-correlated, and adapting the output measurement by using the minus-coordinate projection of $G^{(2)}$. This measurement will have lower contrast and sensitivity than entangled photons and will face issues with camera crosstalk, but could benefit from a higher brightness.} {Finally, it should also be noted that prior studies~\cite{bonato_even_2008,filpi_experimental_2015,simon_spatial_2009,black_quantum_2019,simon_correlated_2011} have explored the use of entangled photon pairs to correct specific types of optical aberrations, but without employing AO.}

{In summary, we have demonstrated that QAO works for bright-field imaging (classical and quantum) and that it can also extend to more complex label-free modalities, such as phase imaging and reflection configurations. Another crucial point is that QAO can be used in all the quantum versions of these systems}~\cite{camphausen_quantumenhanced_2021,black_quantumenhanced_2023,hodgson_reconfigurable_2023,nasr_demonstration_2003,abouraddy_quantumoptical_2002,ndagano_quantum_2022}. This could prove very useful because, as shown in the bright-field case {in Figure~\ref{Figure5} and in a quantum-enhanced phase scheme in supplementary Figure S20}, such quantum schemes are extremely sensitive to optical aberrations, to the point of preventing them from working. QAO {thus has} the potential to optimize the operation of any imaging system based on photon pairs, and could therefore play a major role in the development of future quantum optical microscopes.

\bibliographystyle{apsrev4-2}
\bibliography{biblio}% Produces the bibliography via BibTeX.
% \printbibliography[title={References and Notes}]

 \section*{Acknowledgements}
H.D. and P.C thanks Khaled Kassem, Thomas Chaigne and Amaury Badon for their additional experimental help and discussion. \\
\noindent \textbf{Funding:} D.F. acknowledges support from the Royal Academy
of Engineering Chairs in Emerging Technologies Scheme
and funding from the United Kingdom Engineering
and Physical Sciences Research Council (Grants No.
EP/M01326X/1, EP/R030081/1, EP/Y029097/1) and from the
European Union Horizon 2020 research and innovation program under Grant Agreement No. 801060. H.D. acknowledges funding from the ERC Starting Grant (Grant No. SQIMIC-101039375). P.C and H.D. acknowledge support from SPIE Early Career Researcher Accelerator fund in Quantum Photonics. \\
\noindent \textbf{Author contributions:} P.C. analysed the results and performed the experiment. P.C., B.C., C.V. and R.P. designed the experiments. P.C and H.D. conceived the original ideal. H.D and D.F supervised the project. All authors discussed the data and contributed to the manuscript.\\
\noindent \textbf{Competing interests:} The authors declare that they have no competing interests.\\

\end{document}